\documentclass[showpacs,showkeys,12pt,preprint,preprintnumbers,nofootinbib,
groupedaddress,epsfig,superscriptaddress,amsmath,amssymb]{revtex4}

\usepackage{graphicx}
\usepackage{dcolumn}
\usepackage{bm}
\usepackage{amsmath}
\usepackage{epsfig} 
\usepackage{here}

\def\beq{\begin{equation}}
\def\eeq{\end{equation}}
\def\bea{\begin{array}}
\def\eea{\end{array}}
\def\be{\begin{equation}}
\def\ee{\end{equation}}
\def\ba{\begin{eqnarray}}
\def\ea{\end{eqnarray}}

\def\to{\rightarrow}

\def\[{\left[}
\def\]{\right]}
\def\({\left(}
\def\){\right)}

\def\sm0{{\widetilde{m}_0}}

\def\U1em{{U(1)_{\rm em}}}
\def\to{\rightarrow}

\def\sq2{\sqrt{2}}

\def\ee{e^+e^-}

\def\End{\end{document}}

\def\fsl#1{\setbox0=\hbox{$#1$}                 
   \dimen0=\wd0                                 
   \setbox1=\hbox{/} \dimen1=\wd1               
   \ifdim\dimen0>\dimen1                        
      \rlap{\hbox to \dimen0{\hfil/\hfil}}      
      #1                                        
   \else                                        
      \rlap{\hbox to \dimen1{\hfil$#1$\hfil}}   
      /                                         
   \fi}

\begin{document}

\title{Testing Higgs models via the $H^\pm W^\mp Z$ vertex by a recoil method at the International Linear Collider}
\author{Shinya Kanemura}
\email{kanemu@sci.u-toyama.ac.jp}
\affiliation{Department~of Physics,~University~of~Toyama,~3190~Gofuku,~Toyama~930-8555,~Japan}
\author{Kei Yagyu}
\email{keiyagyu@jodo.sci.u-toyama.ac.jp}
\affiliation{Department~of Physics,~University~of~Toyama,~3190~Gofuku,~Toyama~930-8555,~Japan}
\author{Kazuya Yanase}
\email{yanase@jodo.sci.u-toyama.ac.jp}
\affiliation{Department~of Physics,~University~of~Toyama,~3190~Gofuku,~Toyama~930-8555,~Japan}
%
%
\preprint{UT-HET 050}

\begin{abstract}
In general, charged Higgs bosons $H^\pm$ appear in non-minimal Higgs models. 
The $H^\pm W^\mp Z$ vertex is known to be related to the violation of 
the global symmetry (custodial symmetry) in the Higgs sector. 
Its magnitude strongly depends on the structure of the exotic Higgs models 
which contain higher isospin $SU(2)_L$ representations such as triplet Higgs 
bosons. 
We study the possibility of measuring the $H^\pm W^\mp Z$ vertex 
via single charged Higgs boson production associated with the 
$W^\pm$ boson at the International Linear Collider (ILC) by using the 
recoil method. 
The feasibility of the signal $e^+e^-\to H^\pm W^\mp \to \ell \nu jj$ 
is analyzed assuming the polarized electron and positron beams and  
the expected detector performance for the resolution of the two-jet 
system at the ILC.
The background events can be reduced to a considerable extent 
by imposing the kinematic cuts 
even if we take into account the initial 
state radiation. 
For a relatively light charged Higgs boson whose mass 
$m_{H^\pm}$ is in the region of 120-130 GeV $< m_{H^\pm} < m_W+m_Z$, 
the $H^\pm W^\mp Z$ vertex would be precisely testable especially 
when the decay of $H^\pm$ is lepton specific.
The exoticness of the extended Higgs sector can be explored 
by using combined information for this vertex and the rho parameter.
%
\end{abstract}

\maketitle

\setcounter{footnote}{0}
\renewcommand{\thefootnote}{\arabic{footnote}}

\section{Introduction}

Physics of electroweak symmetry breaking remains unknown, and 
its exploration is crucial to establish our standard picture for 
the origin of masses of elementary particles. 
The structure of the Higgs sector may not necessarily be 
the minimal form in the standard model (SM). 
Extended Higgs sectors have often been considered in various 
new physics contexts beyond the SM.
Therefore, determination of the Higgs sector is 
also important to obtain clue to a new paradigm for 
physics at the TeV scale. 
The Higgs boson search is currently one of the most important tasks 
at the Fermilab Tevatron and the CERN Large Hadron Collider (LHC). 

Basic properties in an extended Higgs sector are the number of 
the scalar fields as well as 
their representation under the isospin $SU(2)_L$ and 
the hypercharge $U(1)_Y$. 
An important observable to constrain the structure of extended 
Higgs models is the electroweak rho parameter $\rho$, 
whose experimental value is very close to unity; $\rho_{\rm exp}=1.0008 
^{+0.0017}_{-0.0007}$~\cite{rho_exp}.  
This fact suggests that a global $SU(2)$ symmetry (custodial symmetry)  
plays an important role in the Higgs sector. 
In the Higgs model which contains complex scalar fields  
with the isospin $T_i$ and the hypercharge $Y_i$ 
as well as real ($Y=0$) scalar fields with the isospin $T_i'$, 
the rho parameter is given at the tree level by    
\begin{align}
\rho_{\textrm{tree}}&=
\frac{\sum_i\left[|v_i|^2(T_i(T_i+1)-Y_i^2)+u_i^2T_i'(T_i'+1)\right]}
{2\sum_i|v_i|^2Y_i^2},   \label{rho_tree}
\end{align}
where $v_i$ ($u_i$) represents the vacuum expectation value (VEV) 
of the complex (real) scalar field~\cite{rho_formula}. 
In the model with only scalar doublet fields (and singlets), 
we obtain $\rho_\textrm{tree}=1$ so that the natural 
extension of the Higgs sector is attained by adding 
extra doublet fields and singlet fields~\cite{rho}. 
In these models, radiative corrections can provide a deviation from 
unity. 
It is well known that in the SM with one Higgs doublet field, 
the mass $m_{H_{SM}}$ of the Higgs boson $H_{SM}$ 
is strongly constrained from above 
through the quantum effect on the rho parameter 
($m_{H_{SM}} < 144$ GeV at 95\% C.L.)~\cite{rho_loop}. 
This bound is clearly model dependent in the non-minimal Higgs sector.
On the other hand, addition of the Higgs field with the isospin 
larger than one half can shift the rho parameter from unity 
at the tree level, whose deviation is proportional to the VEVs of 
these exotic scalar fields. 
The rho parameter, therefore, has been used to exclude or to constrain 
a class of Higgs models~\cite{STU}. 

A common feature in the extended Higgs models 
is the appearance of physical charged scalar components. 
Most of  the extended Higgs models contain 
singly charged Higgs bosons $H^\pm$. 
Hence, we may be able to discriminate each Higgs model through 
the physics of charged Higgs bosons. 
In particular, the $H^\pm W^\mp Z$ vertex can be a useful probe 
of the extended Higgs sector~\cite{Grifols-Mendez,HWZ,HWZ-Kanemura,logan}.  
Assuming that there are several physical charged scalar states 
$H_{\alpha}^\pm$ ($\alpha \geq 2$) and the Nambu-Goldstome modes 
$H_1^\pm$,  The vertex parameter $\xi_\alpha$ in 
$\mathcal{L}=
igm_W \xi_\alpha H_\alpha^+ W^- Z+\textrm{h.c.}$ 
is calculated at the tree level as~\cite{Grifols-Mendez}
\begin{align}
\sum_{\alpha \geq 2} |\xi_\alpha|^2 
&=\frac{1}{\cos^2\theta_W}\left[\frac{2g^2}{m_W^2}
\Big\{\sum_i[T_i(T_i+1)-Y_i^2]|v_i|^2Y_i^2\Big\}-\frac{1}
{\rho_{\textrm{tree}}^2}\right], 
\end{align}
where $\rho_{\rm tree}$ is given in Eq.~(\ref{rho_tree}). 
A non-zero value of $\xi_\alpha$ appears at the tree level 
only when $H_\alpha^\pm$ comes from an exotic representation such as 
triplets. 
Similarly to the case of the rho parameter, 
the vertex is related to the custodial symmetry. 
In general, this can be independent of the rho parameter. 
If a charged Higgs boson $H^\pm_\alpha$ is from a doublet field, 
$\xi_\alpha$ vanishes at the tree level. 
The vertex is then one-loop induced 
and its magnitude is proportional to the violation of the global symmetry 
in the sector of particles in the loop. 
Therefore, the determination of the $H^\pm W^\mp Z$ vertex 
can be a complementary tool to the rho parameter in testing the 
{\it exoticness} of the Higgs sector. 

It goes without saying that 
the decay of $H^\pm$ is strongly model dependent. 
If $H^\pm$ comes from multi-doublet models 
it can couple to quarks and leptons via Yukawa interaction 
which is classified in terms of the softly-broken discrete $Z_2$ 
symmetry to avoid the flavor changing neutral 
current~\cite{Glashow:1976nt,thdm_Yukawa1,thdm_Yukawa2}. 
On the other hand, if $H^\pm$ is originated from 
higher representation fields such as triplets, 
the coupling to quarks is forbidden 
because of the $U(1)_Y$ hypercharge invariance. 
Therefore, the decay of charged Higgs bosons from exotic 
Higgs sectors is mainly leptophilic, 
and they can only couple to quarks through  
mixing with doublet-like charged scalars. 

There have been lots of studies on collider phenomenology 
for charged Higgs bosons. 
At hadron colliders such as the Tevatron and the LHC, 
a main production mechanism 
may be the one from top-quark decay from 
top-quark pair production 
if $m_{H^\pm} < m_t + m_b$ \cite{topdecay}. 
Otherwise, $H^\pm$ may be produced via 
$gg (q\bar{q}) \to tbH^\pm$~\cite{SingleH+},  
$gg (b\bar{b}) \to W^\pm H^\mp$~\cite{ggWH+}, 
$gg (q\bar{q}) \to H^+H^-$~\cite{ggHpHm,qqPairH+}, 
$q\bar{q}'\to W^* \to \phi^0H^\pm$~\cite{AH+}, 
$q\bar{q}'\to W^* \to H^\pm H^{\mp\mp}$~\cite{H-H++}, 
$q \bar q \to q \bar{q}' W^{\pm \ast}Z^\ast\to q \bar{q}' H^\pm$~\cite{WZfusion} 
etc.. 
At the International Linear Collider (ILC)~\cite{ILCTDR2007}, 
$H^\pm$ can be mainly produced in pair 
$e^+e^-\to H^+H^-$~\cite{ILC_PairH+} 
and $\gamma\gamma\to H^+H^-$~\cite{Gamma_PairH+} 
as long as kinematically accessible,  
and if not, single $H^\pm$ production processes may also 
be useful; $e^+e^- \to tbH^\pm$, $e^+e^- \to \tau\nu H^\pm$~\cite{ILC_ffH+}, 
$e^+e^-\to W^\pm H^\mp$~\cite{Cheung:1994rp,eeHW1,eeHW2}, 
$\gamma\gamma (e^- \gamma) \to f \bar{f'} H^\pm$~\cite{Gamma_SingleH+,
eGamma_SingleH+}, etc..  

In this paper, we discuss how accurately the $H^\pm W^\mp Z$ vertex 
can be determined at the collider experiments. 
At the LHC, the vertex would be determined by using the single $H^\pm$ 
production from the $WZ$ fusion\cite{WZfusion}. 
The results are strongly model dependent, 
and the vertex may not be measured unless the $H^\pm$ is light enough 
and $|\xi_\alpha|^2$ is greater than $10^{-2}$. 
If kinematically allowed, the $H^\pm W^\mp Z$ vertex may also 
be measured via the 
decay process of $H^\pm \to W^\pm Z$~\cite{HWZ,HWZ-Kanemura}. 

We here focus on the process $e^+e^- \to W^\pm H^\mp$ 
at the ILC~\cite{Cheung:1994rp,eeHW1,eeHW2}. 
At the ILC, the neutral Higgs boson is produced via 
the Higgs strahlung process $e^+e^-\to ZH$~\cite{ZH}. 
The mass of the Higgs boson can be determined 
in a model independent way by using the so-called recoil method~\cite{recoil}, 
where the information of the Higgs boson can be extracted 
by measuring the leptonic decay products of the recoiled $Z$ boson. 
In this paper we employ this method to test the $H^\pm W^\mp Z$ vertex 
via $e^+e^- \to W^\pm H^\mp$. 
We analyze the signal and backgrounds at the parton level by using 
CalcHEP~\cite{CalcHEP}. 
We take into account the beam polarization and the expected 
resolution for the two-jet system.    
We find that assuming that $H^\pm$ decays into lepton pairs, 
the $H^\pm W^\mp Z$ vertex can be explored accurately by measuring the 
signal of the two-jet with one charged lepton with missing momentum.
For relatively light charged Higgs bosons,   
the signal significance with the value of $|\xi_\alpha|^2 \sim O(10^{-3})$ 
can be as large as two  after appropriate kinematic cuts 
for the collision energy $\sqrt{s} = 300$ GeV and the 
integrated luminosity 1 ab$^{-1}$, 
even when the initial state radiation (ISR) is taken into account.  

This paper is organized as follows. 
We give a quick review for the $H^\pm W^\mp Z$ vertex, and 
discuss the signal process $e^+e^- \to H^\pm W^\mp$ in Sec.~II. 
The feasibility of the signal is analyzed in Sec.~III. 
Some discussions are given in Sec.~IV, and 
conclusions are given in Sec.~V.

\section{The $H^\pm W^\mp Z$ vertex and the process $e^+e^-\to H^\pm W^\mp$}

\subsection{The $H^\pm W^\mp Z$ vertex}

\begin{figure}[t]
\begin{center}
\includegraphics[width=90mm]{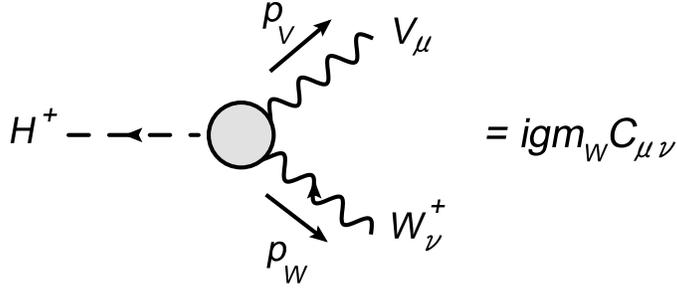}
\caption{The $H^\pm W^\mp V$ vertex ($V=Z$ or $\gamma$).}
\label{hwvf}
\end{center}
\end{figure}
The $H^\pm W^\mp V$ vertex ($V=Z$ or $\gamma$) 
is defined in FIG. \ref{hwvf}, where   
$C^{\mu\nu}$ is expressed in terms of the form factors 
$F_{HWV}$, $G_{HWV}$ and $H_{HWV}$ as 
\begin{align}
C^{\mu\nu}=F_{HWV}g^{\mu\nu}+G_{HWV}\frac{p_W^\mu p_V^\nu}{m_W^2}
       +iH_{HWV}\frac{p_{W\rho} p_{V\sigma}}{m_W^2}\epsilon^{\mu\nu\rho\sigma},
\label{HWV}
\end{align}
with $\epsilon_{\mu\nu\rho\sigma}$ being the anti-symmetric tensor 
with $\epsilon_{0123}=+1$, and $p_V^\mu$ and $p_W^\mu$ being the 
outgoing momenta of $V$ and $W$ bosons, respectively. 
Among the form factors, $F_{HW\gamma}=0$ is derived at the tree level
 due to the $U(1)_{\rm em}$ gauge invariance 
in any extended Higgs models. 
These form factors $F_{HWV}$, $G_{HWV}$ and $H_{HWV}$ 
are respectively related to the coefficients 
$f_{HWV}$, $g_{HWV}$ and $h_{HWV}$ in the effective 
Lagrangian~\cite{HWZ,HWZ-Kanemura}; 
\begin{align}
\mathcal{L}_{\text{eff}}&=gm_Wf_{HWV}H^\pm W_\mu^\mp V^\mu 
                          +g_{HWV}H^\pm F_V^{\mu\nu} F_{W\mu\nu} 
+(ih_{HWV} \epsilon_{\mu\nu\rho\sigma}H^\pm F_V^{\mu\nu} F_W^{\rho\sigma} +\text{H.c.}),
\end{align}
where $F_V^{\mu\nu}$, and $F_W^{\mu\nu}$ are the field strengths. 
We note that $f_{HWZ}$ is the coefficient of the dimension 
three operator, while  
the $g_{HWV}$ and $h_{HWV}$ are those of the dimension five operator,  
so that only $f_{HWZ}$ may appear at the tree level. 
Therefore, the dominant contribution to the $H^\pm W^\mp V$ vertex 
is expected to be from $F_{HWZ}$.  

In the Higgs model with only doublet scalar fields (plus singlets) 
all the form factors including $F_{HWZ}$ vanish 
at the tree level~\cite{Grifols-Mendez}, because of the 
custodial invariance in the kinetic term. 
The form factors  $F_{HWV}$, $G_{HWV}$ and $H_{HWV}$ ($V=\gamma$ and $Z$) 
are generally induced at the loop level. 
In particular, the leading one-loop contribution to $F_{HWZ}$ are 
propotional to the violation of 
the custodial symmetry in the sector of the particle in the loop.
For example, in the two-Higgs-doublet model, the custodial 
symmetry is largely broken via the $t$-$b$ loop contribution as well 
as via the Higgs sector with the mass difference between 
the CP-odd Higgs boson ($A^0$) and the charged Higgs boson 
$H^\pm$~\cite{HWZ-Kanemura}.   
The one-loop induced form factors are theoretically constrained 
from above by perturbative unitarity \cite{Kanemura:1993hm}. 
In such a case, the effect of the custodial symmetry violation also 
can deviate the rho parameter from unity at the one loop level. 
However, when the lightest of CP-even neutral Higgs bosons 
is approximately regarded as  the SM-like Higgs boson, 
the rho parameter can be unity even with a large mass splitting between 
$A^0$ and $H^\pm$  
when the masses of the heavier CP-even neutral Higgs 
boson $H^0$ and $H^\pm$ are common~\cite{Haber-Pomarol}. 
This means that the appearance of the 
$H^\pm W^\mp Z$ vertex and the deviation from unity in the rho parameter 
are not necessarily correlated at the one-loop level, 
and they can be independent quantities, in principle.

The simplest models in which the $H^\pm W^\mp Z$ vertex appears 
at the tree level are those with triplet scalar fields.
In the model with an isospin doublet field ($Y=1/2$) 
and either an real triplet field $\eta$ ($Y=0$) 
or an additional complex triplet field $\Delta$ ($Y=1$),  
concrete expressions for the tree-level formulae for  
$|F_{HWZ}|^2$ and that of $\rho_{\textrm{tree}}$ are 
shown in TABLE~\ref{f_models}, 
where $v$, $v_\eta$ and $v_\Delta$  are respectively 
VEVs of the doublet scalar field and 
the additional triplet scalar field $\eta$ and $\Delta$. 
These triplet scalar fields also contribute to 
the rho parameter at the tree level, so that their VEVs 
are constrained by the current rho parameter data,  
$\rho_{\exp}=1.0008^{+0.0017}_{-0.0007}$; i.e., 
$v_\eta \lesssim 6$ GeV for the real triplet field $\eta$, 
and $v_\Delta \lesssim 8$ GeV for the complex triplet $\Delta$ 
(95 \% CL). 
We note that in order to obtain the similar accuracy to the rho 
parameter data by measuring the $H^\pm W^\mp Z$ vertex, 
the vertex  has to be measured with the detectability to    
$|F_{HWZ}|^2 \sim {\cal O}(10^{-3})$.

\begin{table}[t] 
\begin{center}
{\renewcommand\arraystretch{1.2}
\begin{tabular}{|c||c|c|c|}\hline
Model & SM with $\eta$  ($Y=0$)& SM with $\Delta$ ($Y=1$) 
& the GM model \\\hline\hline
$|F_{HWZ}|^2=$&$\frac{4v^2 v^{2}_\eta}{\cos^2\theta_W(v^2+4v^{2}_\eta)^2}$
&$\frac{2v^2 v^{2}_\Delta}{\cos^2\theta_W(v^2+2v^{2}_\Delta)^2 }$& 
$\frac{4 v_\Delta^2}{\cos^2\theta_W(v^2+4v_\Delta^2)}$ 
\\\hline 
$\rho_{\text{tree}}=$&$1+\frac{4v^{2}_\eta}{v^2}$
&$\frac{1+2\frac{v^{2}_\Delta}{v^2}}{1+4\frac{v^{2}_\Delta}{v^2}}$&$1$ \\\hline
\end{tabular}}
\caption{The tree-level expression 
for $F_{HWZ}$ and rho parameter at the tree level 
in the model with a real triplet field, that with a complex triplet field 
and the Georgi-Machacek (GM) model~\cite{Georgi:1985nv}.}
\label{f_models}
\end{center}
\end{table}

Finally, we mention about the model with a real triplet field $\eta$ 
and a complex triplet field $\Delta$ in addition to the SM, 
which is proposed by Georgi-Machacek and 
Chanowiz-Golden~\cite{Georgi:1985nv, Gunion:1989ci,AK_GM}.
In this model, an alignment of the VEVs for $\eta$ and $\Delta$ 
are introduced ($v_\eta= v_\Delta/\sqrt{2}$), 
by which the Higgs potential is invariant under   
the custodial $SU(2)$ symmetry at the tree level. 
Physical scalar states in this model can be classified 
using the transformation property against 
the custodial symmetry; i.e., 
the five-plet, the three-plet and the singlet. 
Only the charged Higgs boson from the five-plet state has 
the non-zero value of $F_{HWZ}$ at the tree level. 
Its value is proportional to the VEV $v_\Delta$ for the triplet scalar 
fields. 
However, the value of $v_\Delta$ is not strongly constrained 
by the rho parameter data, because 
the tree level contribution to the rho parameter 
is zero due to the custodial symmetry: 
see TABLE~\ref{f_models}. 
Consequently, the magnitude of $|F_{HWZ}|^2$ 
can be of order one.

\subsection{The $e^+e^-\to H^\pm W^\mp$ process}

\begin{figure}[t]
\begin{center}
\includegraphics[width=70mm]{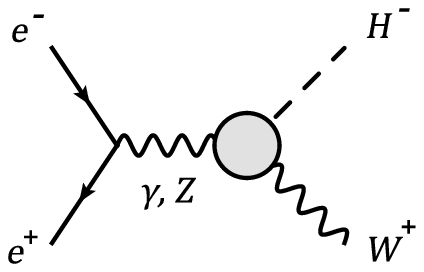}
\caption{The $e^+e^- \to H^-W^+$ process}
\label{eehwa}
\end{center}
\end{figure}

The process $e^+ e^-\to H^-   W^+$ is depicted in FIG.~\ref{eehwa}. 
This process is directly related to the $H^\pm W^\mp Z$ vertex. 
The helicity amplitudes are calculated by 
\begin{align}
\mathcal{M}(\tau,\lambda)
&=\sum_{V=Z,\gamma}igm_WC_V\frac{1}{s-m_V^2}j_\mu(\tau) C^{\mu\nu}
\epsilon_\nu(\lambda),
\end{align}
where $\sqrt{s}$ is the center-of-mass energy, 
$j_\mu(\tau)$ is the electron current, and 
$\epsilon_\nu(\lambda)$ is the polarization vector of the $W^+$ 
boson~\cite{eeHW1}. 
The helicities of the electron and the $W^+$ boson 
can be $\tau=\pm 1$ and $\lambda=0,\pm 1$, respectively.  
The coefficient $C_V$ is given by
\begin{align}
C_V=\left\{
\begin{array}{cc}
eQ_e,&\hspace{4mm}\text{for }V=\gamma,\\
\frac{g}{\cos\theta_W}(T_e^3-\sin^2\theta_WQ_e),
&\hspace{4mm}\text{for }V=Z,\label{c}
\end{array}
\right.
\end{align}
with $Q_e=-1$, $T_e^3=-1/2$ $(0)$ for $\tau=-1$ ($+1$). 
The squared amplitude is evaluated as
\begin{align}
&|\mathcal{M}(\tau)|^2\equiv
\sum_{\lambda=0,\pm}|\mathcal{M}(\tau,\lambda)|^2\notag\\
&=g^2\left|C_\gamma\frac{F_{HW\gamma}}{s}+
C_Z\frac{F_{HWZ}}{s-m_Z^2}\right|^2\left[\frac{\sin^2\theta}{4}
(s+m_W^2-m_{H^\pm}^2)^2+sm_W^2(\cos^2\theta+1)\right],\label{amp2}
\end{align}
where $\theta$ is the angle between the momentum of $H^\pm$ and the 
beam axis, $m_{H^\pm}$ is the mass of $H^\pm$ 
and the form factors $G_{HWV}$ and $H_{HWV}$ are taken to be zero. 
The helicity specified cross sections are written 
in terms of the squared amplitude in Eq. (\ref{amp2}), 
\begin{align}
\sigma(s; \tau)=
\frac{1}{32\pi s}\beta\left(\frac{m_{H^\pm}^2}{s},\frac{m_W^2}{s}\right)
\int_{-1}^{1}d\cos\theta|\mathcal{M}(\tau)|^2,
\end{align}
where $\sigma(s; +1)=\sigma(e^+_Le^-_R\to H^-W^+)$ and 
$\sigma(s; -1)=\sigma(e^+_Re^-_L\to H^-W^+)$, and 
\begin{align}
\beta\left(x,y\right)=\sqrt{1+x^2+y^2-2xy-2x-2y}.
\end{align}
The helicity averaged cross section is given by 
$\sigma(e^+e^-\rightarrow H^-W^+)=(\sigma(s, +1)+\sigma(s, -1))/4$.

\begin{figure}[t]
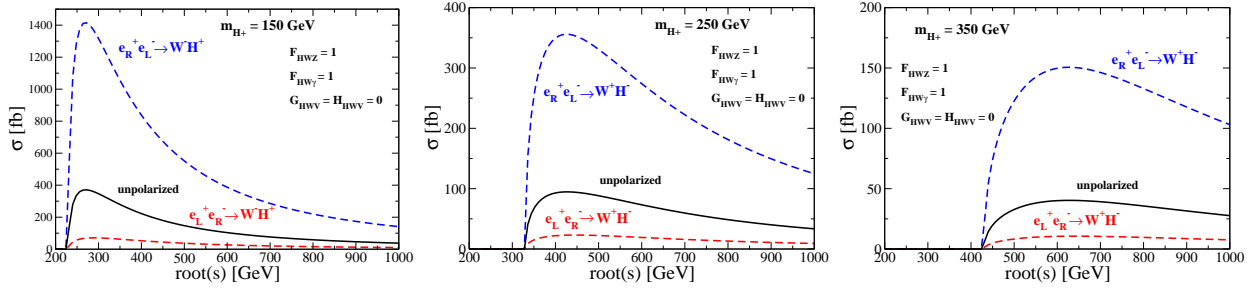

\begin{center}
\includegraphics[width=54mm]{cs_fz1fg1_mch150_v2.eps}
\includegraphics[width=54mm]{cs_fz1fg1_mch250_v2.eps}
\includegraphics[width=54mm]{cs_fz1fg1_mch350_v2.eps}
\caption{The total cross section as a function of $\sqrt{s}$ in the case 
of $F_{HWZ}=F_{HW\gamma}=1$.} \label{mch150_fz1_0}
\end{center}
\end{figure}

\begin{figure}[t]
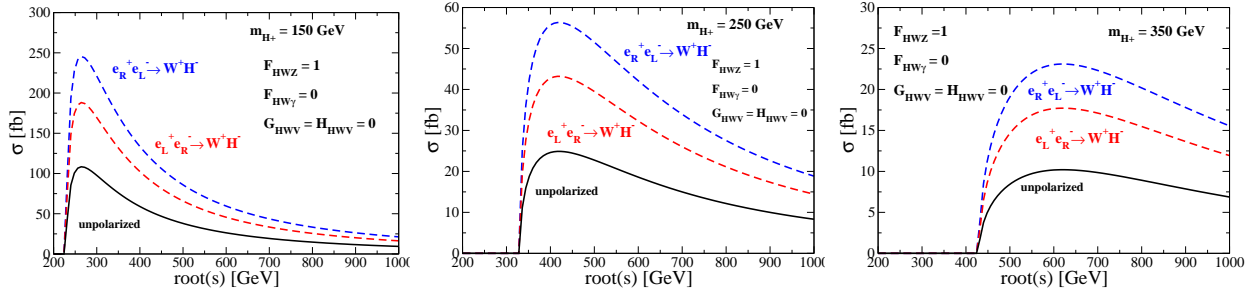

\begin{center}
\includegraphics[width=54mm]{cs_fz1fg0_mch150.eps}
\includegraphics[width=54mm]{cs_fz1fg0_mch250.eps}
\includegraphics[width=54mm]{cs_fz1fg0_mch350.eps}
\caption{The total cross section as a function of $\sqrt{s}$ 
in the case of $F_{HWZ}=1, F_{HW\gamma}=0$.}
\label{mch150_fz1_02}
\end{center}
\end{figure}

In FIGs. \ref{mch150_fz1_0} and \ref{mch150_fz1_02}, we show that the 
$\sqrt{s}$ dependence
 of the helicity dependent and the helicity averaged cross sections. 
Notice that the behavior of these cross sections drastically changes 
depending on the initial electron helicity 
in the case of $F_{HWZ}\simeq F_{HW\gamma}$. 
On the contrary, there is no such a difference 
in the case of $F_{HWZ}\gg F_{HW\gamma}$. 
As mentioned before, $F_{HW\gamma}$ is zero at the tree level in any models 
because of the $U(1)_{\textrm{em}}$ gauge invariance. 
The relation of $F_{HWZ}\gg F_{HW\gamma}$ or $F_{HWZ}\simeq F_{HW\gamma}$ 
can be tested by using the initial electron helicities.  



\section{The signal and background analysis}

\subsection{Recoil method and the assumption for the ILC performance}

We investigate the possibility of measuring 
the $H^\pm W^\mp Z$ vertex by using a recoil method at the ILC. 
It has been known that this method is a useful tool 
for measuring the mass of the SM-like Higgs boson $H_{\rm SM}$  
without assuming the decay branching fraction of the Higgs 
boson~\cite{recoil}. 
In the Higgs-strahlung process $e^+e^- \to ZH_{\text{SM}} $~\cite{ZH}, 
the Higgs boson mass can be obtained as the recoil mass $m_{\text{recoil}}$ 
from two leptons produced from the $Z$ boson, whose 
energy is $E_{\ell\ell}$, and the invariant mass is $M_{\ell\ell}$. 
They satisfy the relation, 
\begin{align}
m_{\text{recoil}}^2(\ell\ell)=s-2\sqrt{s}E_{\ell\ell}+M_{\ell\ell}^2.
\end{align}
The information of the Higgs boson mass can be extracted by 
measuring $E_{\ell\ell}$ and $m_{\ell\ell}$ 
in a model independent way. 

\begin{figure}[t]
\begin{center}
\includegraphics[width=70mm]{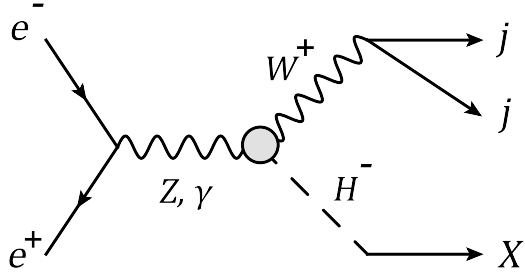}
\caption{The signal process.}
\label{ee_jjX}
\end{center}
\end{figure}

In this paper we apply this method to $e^+e^-\to W^\pm H^\mp $ in order to  
measure the $H^\pm W^\mp Z$ vertex. 
In order to identify the process, 
we consider the hadronic decays $W\to jj$ 
instead of the leptonic decay of the produced $W$ boson, 
and obtain information of the $H^\pm W^\mp Z$ vertex by using the 
recoil of the two-jet system. 
The recoiled mass of $H^\pm$ is given in terms of the 
two-jet energy $E_{jj}$ and 
the two-jet invariant mass $M_{jj}$ as
\begin{align}
m_{\text{recoil}}^2(jj)=s-2\sqrt{s}E_{jj}+M_{jj}^2.\label{recoil2}
\end{align}
This process is shown in FIG.~\ref{ee_jjX}.
It is clear that the detector performance for 
the resolution of two jets is crucial in such an analysis. 
In particular, the jets from the $W$ boson in the signal process 
has to be precisely measured in order to be separated with those from the 
$Z$ boson in the background process. 
At the ILC, the resolution for the two jet system with the energy $E$ 
in the unit of GeV is expected to be $\sigma_E = 0.3 \sqrt{E}$ GeV, 
by which the background 
from $Z\to jj$ can be considerably reduced.  
We here adopt the similar value for $\sigma_E$ ($\sim 3$ GeV) 
in our later analysis. 

At the ILC, the polarized electron and positron beams can be used, by 
which the background from the $W$ boson pair production process can be 
reduced. We here use the following beams polarized as  
\begin{align}
\frac{N_{e_R^-}-N_{e_L^-}}{N_{e_L^-}+N_{e_R^-}} =0.8,\quad 
\frac{N_{e_L^+}-N_{e_R^+}}{N_{e_L^+}+N_{e_R^+}} =0.5,\label{porl}
\end{align}
which are expected to be attained at the ILC~\cite{ILCTDR2007}, 
where 
$N_{e_{R,L}^-}$ and $N_{e_{R,L}^+}$ are numbers of right- (left-) handed electron and 
positron in the beam flux per unit time. 
The total cross sections for the  signal and the backgrounds can be evaluated 
from the helicity specified cross sections as  
\begin{align}
\sigma_{\text{tot}} (e^+e^-\to X)=&x_-x_+\sigma(e_L^+e_R^- \to X)+(1-x_-)(1-x_+) 
\sigma(e_R^+e_L^- \to X)\notag\\
&+x_-(1-x_+)\sigma(e_R^+e_R^- \to X)+x_+(1-x_-)\sigma(e_L^+e_L^- \to X),
\end{align}
where $x_- = N_{e_R^-}/(N_{e_L^-}+N_{e_R^-})$ and 
      $x_+ = N_{e_L^+}/(N_{e_L^+}+N_{e_R^+})$.  

The high-energy electron and positron beams lose their incident energies 
by the ISR. 
In our analysis, we also take into account such effect, and see how the 
results without the ISR are changed by including the effect of the ISR. 

\subsection{Signal and Backgrounds}

The size of the signal cross section is determined by 
the center of mass energy $\sqrt{s}$, the mass $m_{H^\pm}^{}$ and 
the form factors $F_{HWZ}$ and $F_{HW\gamma}$. 
In the following analysis, we consider the case of 
$(F_{HWZ}, F_{HW\gamma}) \equiv (\xi,0)$. 
This approximately corresponds to most of the cases we are interested, such as 
the triplet models.   
In order to examine the possibility of constraining $|\xi|^2$, 
we here assume that the mass of the charged Higgs boson is already 
known with some accuracy 
by measuring the other processes at the LHC or at the ILC. 
Then $|\xi|^2$ is a unique free parameter in the production cross section.  

\begin{figure}[t]
\begin{center}
\includegraphics[width=120mm]{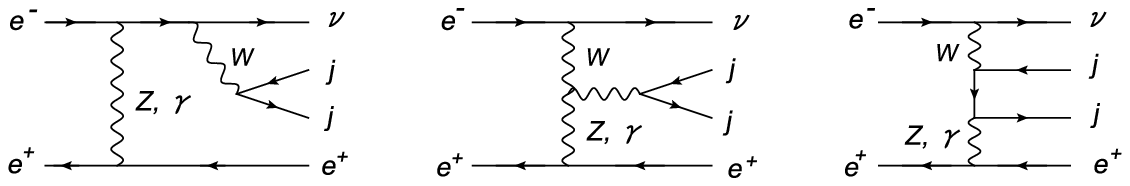}\\
\vspace{5mm}
\includegraphics[width=80mm]{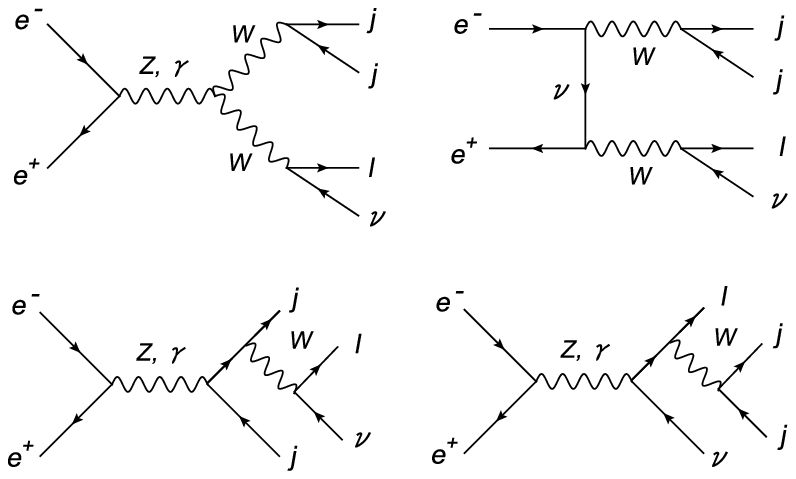}
\caption{The $e^+e^-\to \ell\nu jj$ backgrounds.}
\label{enjj_BG}
\end{center}
\end{figure}

\begin{figure}[t]
\begin{center}
\includegraphics[width=80mm]{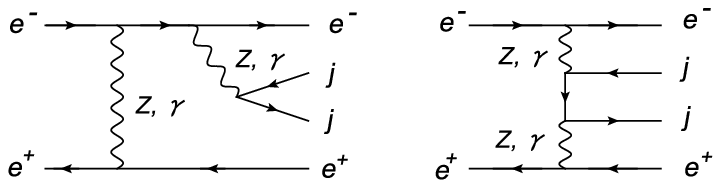}\\
\vspace{5mm}
\includegraphics[width=120mm]{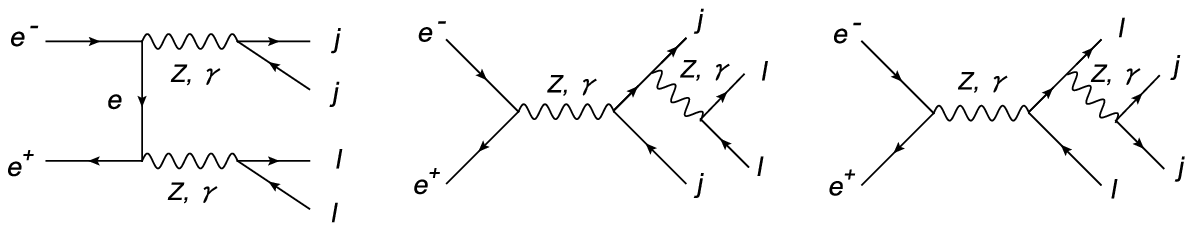}
\caption{The $e^+e^-\to \ell\ell jj$ backgrounds.}
\label{eejj_BG}
\end{center}
\end{figure}

In order to perform the signal and background analysis, 
we here assume that the decay of the produced charged Higgs boson is 
lepton specific; i.e., $H^\pm \to \ell \nu$ where $\ell$ is either 
$e$, $\mu$ or $\tau$.  
The final state of the signal is then $e^+e^- \to H^\pm W^\mp \to \ell \nu jj$. 
We first consider $m_{H^\pm} < m_W+m_Z$ to avoid the complexness 
with the possible decay mode of $H^\pm \to W^\pm Z$, whose 
branching ratio strongly depends on the model. 
The main backgrounds come from the $W$ boson pair production process 
$e^+e^- \to W^+W^-$ and the single $W$ production processes 
in FIG.~\ref{enjj_BG}. 
For the $e^\pm \nu jj$ final state, additional processes 
shown in FIG.~\ref{enjj_BG} (upper figures) 
can also be a significant background. 
In addition, we take into account the processes with the final 
state of $\ell\ell jj$ shown in FIG.~\ref{eejj_BG}. They can be 
backgrounds if one of the outgoing leptons escapes from 
the detection at the detector. We here assume that the miss identity 
rate for a lepton is 10~\%.

We impose the basic cuts for all events such as 
\begin{align}
10^\circ < A_j < 170^\circ,\quad 5^\circ < A_{jj} < 175^\circ,
\quad 10\; \textrm{GeV}< E_{jj}, \label{basic}
\end{align}
where $A_j$ is the angle between a jet and the beam axis, 
$A_{jj}$ is the angle between the two jets 
and $E_{jj}$ is the energy of the two jets. 
In the numerical evaluation, we use CalcHEP~\cite{CalcHEP}.

After the basic cuts, the event numbers of both the signal and the 
backgrounds are listed in TABLE~\ref{result1} for the case without ISR, and 
in TABLE~\ref{result2} for that with ISR, 
where the center of mass energy is set $\sqrt{s}=300$ GeV, the mass of the charged Higgs boson $m_{H^\pm}$ is 150 GeV,  
and the parameter $|\xi|^2$ for the $H^\pm W^\mp Z$ vertex is set to be $10^{-3}$. 
For both the cases signal over background 
ratios are less than $10^{-4}$ before 
imposing the other kinematic cuts than the basic cuts in Eq.~(\ref{basic}). 
In the following we first discuss the case without the ISR, 
then present the results for that with the ISR.

\begin{table}[!h]
\begin{center}
{\renewcommand\arraystretch{1.3}
\begin{tabular}{|l||c|c|c|c||c|c|}\hline
 &Basic & $M_{jj}$ \hspace{5mm}  &  $p_T^{jj}$\hspace{5mm}  & $E_{jj}$ \hspace{5mm} & $\cos\theta_{\text{lep}}$ & $M_{\ell \nu}$\hspace{5mm}  \\\hline\hline
$e_R^+e_L^-\to \ell^\pm \nu jj$~(fb) & 7.2$\times 10^{-3}$ &6.4$\times 10^{-3}$ &4.4$\times 10^{-3}$  &4.4$\times 10^{-3}$  &3.3$\times 10^{-3}$ &3.3$\times 10^{-3}$\\\hline
$e_L^+e_R^-\to \ell^\pm \nu jj$~(fb) & 1.4$\times 10^{-1}$&1.3$\times 10^{-1}$ &8.5$\times 10^{-2}$  &8.5$\times 10^{-2}$ &6.7$\times 10^{-2}$ &6.7$\times 10^{-2}$ \\\hline\hline
Total signal~(fb) & 1.5$\times 10^{-1}$ &1.4$\times 10^{-1}$&8.9$\times 10^{-2}$  &8.9$\times 10^{-2}$ &7.0$\times 10^{-2}$&7.0$\times 10^{-2}$\\\hline\hline
$e_R^+e_L^-\to \mu^\pm \nu jj+\tau^\pm \nu jj$~(fb) &340 &300  &53  &2.9$\times 10^{-1}$ &2.2$\times 10^{-1}$&1.3$\times 10^{-1}$\\\hline
$e_L^+e_R^-\to \mu^\pm \nu jj+\tau^\pm \nu jj$~(fb) &80 &71 &13  &2.8$\times 10^{-1}$ &2.1$\times 10^{-1}$&1.1$\times 10^{-1}$\\\hline
$e_R^+e_L^-\to e^\pm \nu jj$~(fb) &220 &190 &31 &1.6 &6.4$\times 10^{-1}$&3.4$\times 10^{-1}$\\\hline
$e_L^+e_R^-\to e^\pm \nu jj$~(fb) &40 &36 &6.4 &1.4$\times 10^{-1}$ &1.1$\times 10^{-1}$&5.7$\times 10^{-2}$\\\hline
$e_R^+e_R^-\to e_R^- \bar{\nu} jj$~(fb) &100 &92 &11 &3.8 &2.2$\times 10^{-1}$&1.2$\times 10^{-1}$\\\hline
$e_L^+e_L^-\to e_L^+ \nu jj$~(fb) &40 &31 &4.3 &1.3 &7.2$\times 10^{-2}$&4.1$\times 10^{-2}$\\\hline\hline
Total $\ell\nu jj$ background~(fb)&820 &720 &120 &7.4 &1.5 &8.0$\times 10^{-1}$\\\hline\hline
$e_R^+e_L^-\to \mu^+ \mu^- jj+\tau^+ \tau^- jj$~(fb) &1.2 &3.7$\times 10^{-2}$ &5.5$\times 10^{-3}$ &1.1$\times 10^{-4}$ &9.4$\times 10^{-5}$ &5.0$\times 10^{-5}$\\\hline
$e_L^+e_R^-\to \mu^+ \mu^- jj+\tau^+ \tau^- jj$~(fb) &19 &1.0  &1.4$\times 10^{-1}$ &3.0$\times 10^{-3}$ &2.5$\times 10^{-3}$ &1.4$\times 10^{-3}$\\\hline
$e_R^+e_L^-\to e^+ e^- jj$~(fb) &8.4 &9.0$\times 10^{-2}$ &4.6$\times 10^{-3}$ &5.8$\times 10^{-4}$ &2.6$\times 10^{-4}$ &1.3$\times 10^{-4}$ \\\hline
$e_L^+e_R^-\to e^+ e^- jj$~(fb) &220 &2.4 &1.2$\times 10^{-1}$ &1.5$\times 10^{-2}$ &6.7$\times 10^{-3}$ &3.4$\times 10^{-3}$ \\\hline
$e_R^+e_R^-\to e^+ e^- jj$~(fb) &59 &7.2$\times 10^{-1}$ &2.4$\times 10^{-2}$ &4.5$\times 10^{-3}$ &2.0$\times 10^{-3}$ &1.0$\times 10^{-3}$ \\\hline
$e_L^+e_L^-\to e^+ e^- jj$~(fb) &19 &1.0  &8.0$\times 10^{-3}$ &1.4$\times 10^{-3}$ &6.7$\times 10^{-4}$ &3.7$\times 10^{-4}$ \\\hline\hline
Total $\ell\ell jj$ background~(fb) &330 &5.2 &3.0$\times 10^{-1}$ &2.5$\times 10^{-2}$  &1.2$\times 10^{-2}$ &6.4$\times 10^{-3}$\\\hline\hline
$S/\sqrt{B}$ (assuming 1 ab$^{-1}$) &1.4$\times 10^{-1}$&1.6$\times 10^{-1}$&2.6$\times 10^{-1}$&1.0&1.8&2.5\\\hline
$S/B$ (assuming 1 ab$^{-1}$) &1.3$\times 10^{-4}$&1.9$\times 10^{-4}$&7.4$\times 10^{-4}$&1.2$\times 10^{-2}$&4.6$\times 10^{-2}$&8.7$\times 10^{-2}$\\\hline
\end{tabular}}
\caption{The results without ISR. 
The cross sections of both the signal and the backgrounds are 
shown for $\sqrt{s}=300$ GeV.  
For the signal, $m_{H^\pm}$ is 150 GeV and $|\xi|^2$ 
is taken to be 10$^{-3}$. 
For the $\ell\ell jj$ processes, the misidentity rate of one of 
the leptons is assumed to be 0.1. 
The signal significance $S/\sqrt{B}$ and the ratio $S/B$ are evaluated for 
the integrated luminosity to be 1 ab$^{-1}$.}
\label{result1}
\end{center}
\end{table}

\begin{table}[!h]
\begin{center}
{\renewcommand\arraystretch{1.3}
\begin{tabular}{|l||c|c|c|c||c|c|}\hline
 &Basic & $M_{jj}$ \hspace{5mm}  &  $p_T^{jj}$\hspace{5mm}  & $E_{jj}$ \hspace{5mm} & $\cos\theta_{\text{lep}}$ & $M_{\ell \nu}$\hspace{5mm}  \\\hline\hline
$e_R^+e_L^-\to \ell^\pm \nu jj$~(fb) &6.8$\times 10^{-3}$ &6.0$\times 10^{-3}$ &3.3$\times 10^{-3}$ &3.1$\times 10^{-3}$ &2.4$\times 10^{-3}$ &2.4$\times 10^{-3}$\\\hline
$e_L^+e_R^-\to \ell^\pm \nu jj$~(fb) &1.3$\times 10^{-1}$ &1.2$\times 10^{-1}$ &6.6$\times 10^{-2}$ &6.3$\times 10^{-2}$ &5.0$\times 10^{-2}$ &4.9$\times 10^{-2}$\\\hline\hline
Total signal~(fb) &1.4$\times 10^{-1}$ &1.3$\times 10^{-1}$ &6.9$\times 10^{-2}$ &6.6$\times 10^{-2}$ &5.2$\times 10^{-2}$ &5.1$\times 10^{-2}$ \\\hline\hline
$e_R^+e_L^-\to \mu^\pm \nu jj+\tau^\pm \nu jj$~(fb) &350 &310 &55 &2.9 &2.2 &1.1$\times 10^{-1}$\\\hline
$e_L^+e_R^-\to \mu^\pm \nu jj+\tau^\pm \nu jj$~(fb) &84 &76 &17 &1.8 &1.4 &9.7$\times 10^{-2}$\\\hline
$e_R^+e_L^-\to e^\pm \nu jj$~(fb) &210 &190 &32 &2.8 &1.6 &2.8$\times 10^{-1}$ \\\hline
$e_L^+e_R^-\to e^\pm \nu jj$~(fb) &42 &38 &8.5 &9.0$\times 10^{-1}$ &7.0$\times 10^{-1}$ &4.9$\times 10^{-2}$\\\hline
$e_R^+e_R^-\to e_R^- \bar{\nu} jj$~(fb) &92 &81 &10 &3.2 &2.2$\times 10^{-1}$ &1.0$\times 10^{-1}$\\\hline
$e_L^+e_L^-\to e_L^+ \nu jj$~(fb) &32 &29 &3.7 &1.1 &7.8$\times 10^{-2}$ &3.4$\times 10^{-2}$\\\hline\hline
Total $\ell\nu jj$ background~(fb)  &810 &720 &130 &13 &6.2 &6.7$\times 10^{-1}$ \\\hline\hline
$e_R^+e_L^-\to \mu^+ \mu^- jj+\tau^+ \tau^- jj$~(fb) &1.2 &4.2$\times 10^{-2}$ &5.9$\times 10^{-3}$ &3.7$\times 10^{-4}$ &3.1$\times 10^{-4}$&4.6$\times 10^{-5}$ \\\hline
$e_L^+e_R^-\to \mu^+ \mu^- jj+\tau^+ \tau^- jj$~(fb) &22 &1.2 &1.5$\times 10^{-1}$ &9.9$\times 10^{-3}$ &8.3$\times 10^{-3}$ &1.2$\times 10^{-3}$ \\\hline
$e_R^+e_L^-\to e^+ e^- jj$~(fb) &9.6 &9.2$\times 10^{-2}$ &4.1$\times 10^{-3}$ &6.3$\times 10^{-4}$ &3.2$\times 10^{-4}$ &1.0$\times 10^{-4}$\\\hline
$e_L^+e_R^-\to e^+ e^- jj$~(fb) &230 &2.4 &1.0$\times 10^{-1}$ &1.7$\times 10^{-2}$ &9.2$\times 10^{-3}$ &2.9$\times 10^{-3}$ \\\hline
$e_R^+e_R^-\to e^+ e^- jj$~(fb) &70 &6.4$\times 10^{-1}$ &2.3$\times 10^{-2}$ &4.2$\times 10^{-3}$ &2.1$\times 10^{-3}$ &9.1$\times 10^{-4}$\\\hline
$e_L^+e_L^-\to e^+ e^- jj$~(fb) &24 &2.2$\times 10^{-1}$ &7.4$\times 10^{-3}$ &1.4$\times 10^{-3}$ &6.3$\times 10^{-4}$ &3.1$\times 10^{-4}$\\\hline\hline
Total $\ell\ell jj$ background~(fb) &360 &4.6 &2.9$\times 10^{-1}$ &3.4$\times 10^{-2}$ &2.1$\times 10^{-2}$ &5.5$\times 10^{-3}$\\\hline\hline
$S/\sqrt{B}$ (assuming 1 ab$^{-1}$)&1.3$\times 10^{-1}$&1.5$\times 10^{-1}$&1.9$\times 10^{-1}$&5.8$\times 10^{-1}$&6.6$\times 10^{-1}$&2.0\\\hline
$S/B$ (assuming 1 ab$^{-1}$)&1.2$\times 10^{-4}$&1.8$\times 10^{-4}$&5.3$\times 10^{-4}$&5.1$\times 10^{-3}$&8.4$\times 10^{-3}$&7.5$\times 10^{-2}$\\\hline
\end{tabular}}
\caption{The results with the ISR. 
The cross sections of both the signal and the backgrounds are shown for $\sqrt{s}=300$ GeV.  
For the signal, $m_{H^\pm}$ is 150 GeV and  
$|\xi|^2$ is taken to be 10$^{-3}$. 
For the $\ell\ell jj$ processes, the misidentity rate of one 
of the leptons is assumed to be 0.1. 
The signal significance $S/\sqrt{B}$ and the ratio $S/B$ are evaluated for 
the integrated luminosity to be 1 ab$^{-1}$.}
\label{result2}
\end{center}
\end{table}

In order to improve the signal over background 
ratio, we impose additional kinematic cuts.
The two jets come from the $W$ boson for the signal, so that 
the invariant mass cut is useful to reduce the backgrounds where 
a parent of the two jets is not the $W$ boson. 
We here impose the condition; 
\begin{align}
m_W-n\sigma_E < M_{jj} < m_W+n\sigma_E,  \label{mjj}   
\end{align}
where $\sigma_E$ represents the resolution of the detector 
which we assume 3 GeV, and $n$ is taken to be 2 here.

\begin{figure}[t]
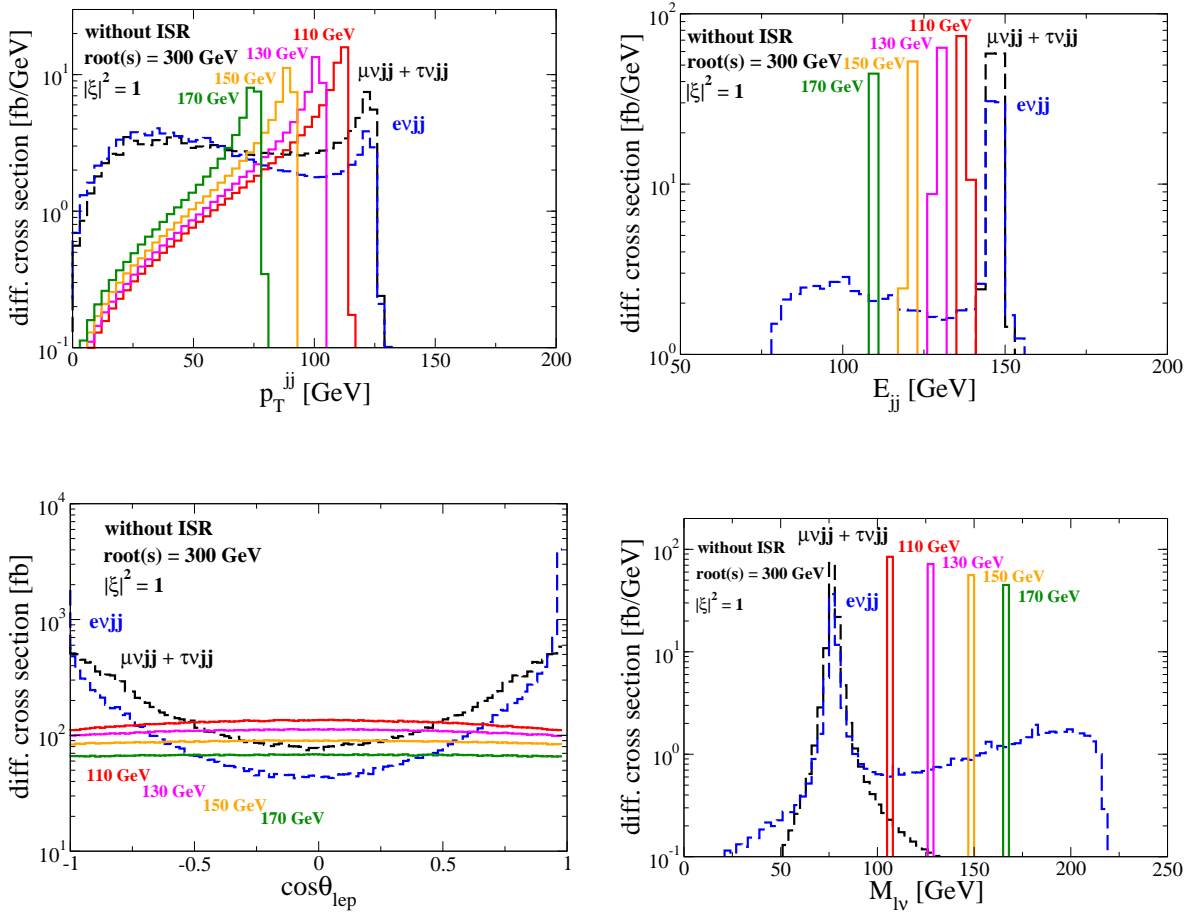

\begin{center}
\includegraphics[width=75mm]{distr_tjj.eps}\hspace{5mm}
\includegraphics[width=75mm]{distr_ejj.eps}\\
\vspace{10mm}
\includegraphics[width=75mm]{distr_clep.eps}\hspace{5mm}
\includegraphics[width=75mm]{distr_mll.eps}
\caption{Distributions of the signal 
for $m_{H^\pm}=110$, 130, 150 and 170 GeV as well as the backgrounds 
after the invariant mass $M_{jj}$ cut in Eq.~(\ref{mjj}) without the ISR   
as a function of the transverse momentum 
$p_T^{jj}$ (upper left), the energy of the $jj$ system (upper right), 
the angle $\theta_{\rm lep}$ of a charged lepton with the beam 
axis (lower left), and 
the invariant mass $M_{\ell\nu}$ of the charged lepton and the missing momentum 
in the final state (lower right).  
$|\xi|^2$ is taken to be 1.}
\label{distr1}
\end{center}
\end{figure}
\begin{figure}[t]
\begin{center}
\includegraphics[width=80mm]{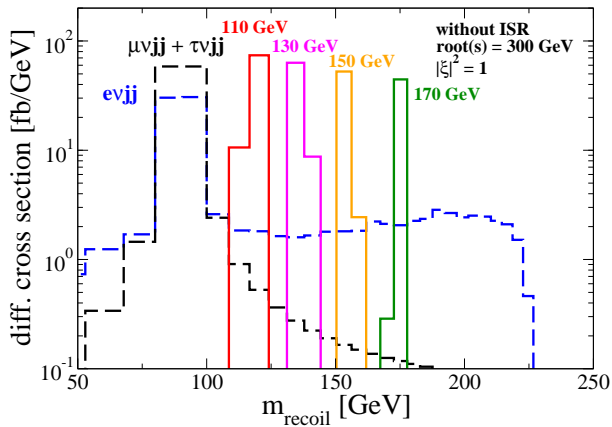}
\caption{Distributions of the signal for $m_{H^\pm}=110$, 130, 150 and 170 GeV
 as well as the backgrounds after the cut in Eq.~(\ref{mjj}) 
without the ISR as a function of the recoil mass $m_{\rm recoil}$.}
\label{distr_recoil}
\end{center}
\end{figure}

In FIG.~\ref{distr1}, the differential cross sections of the signal and the 
backgrounds are shown for the events after the $M_{jj}$ cut in Eq.~(\ref{mjj}) 
as a function of the transverse momentum $p_T^{jj}$, 
the energy of the $jj$ system, the angle $\theta_{\rm lep}$ of a charged 
lepton with the beam axis, and 
the invariant mass $M_{\ell\nu}$ of the charged lepton and the missing momentum 
in the final state.  
For the signal, the results are shown 
for $|\xi|^2=1$ with the mass of the charged Higgs boson to be 
110, 130, 150 and 170 GeV. 
The $E_{jj}$ distribution shown in FIG.~\ref{distr1} (upper-right) can be 
translated into the distribution as a function of $m_{\rm recoil}$ by using the 
relation in Eq.~(\ref{recoil2}), which is shown in FIG.~\ref{distr_recoil}. 
The signal events form the peak at $m_{\rm recoil} \sim m_{H^\pm}$. 

According to FIG.~\ref{distr1}, we impose the following four kinematic cuts sequentially: 
\begin{align}
  75\text{ GeV} <p_T^{jj}<100\text{ GeV}, 
\end{align}
and 
\begin{align}
  115\text{ GeV} <E_{jj}<125\text{ GeV}
\end{align}
for the $jj$ system in the final state.  
In TABLE~\ref{result1}, the resulting values for the cross sections 
for the signal and backgrounds are shown in each step of the cuts.
The backgrounds can be reduced in a considerable extent. 
For $|\xi|^2=10^{-3}$, the signal significance reaches 
to ${\cal O}(1)$ assuming the integrated luminosity of 
1 ab$^{-1}$.  

Until now, we have imposed the cuts on the $jj$ system, 
and no information from the $\ell \nu$ system has been used.  
Here, in order to further improve the signal significance,   
we impose new cuts related to the $\ell \nu$ system in order, which 
are determined from FIG.~\ref{distr1}; 
\begin{align}
|\cos\theta_{\text{lep}}| < 0.75,
\end{align}
and 
\begin{align}
144\text{ GeV} <M_{\ell\nu}<156\text{ GeV}. 
\end{align}
As shown in TABLE~\ref{result1}, for $|\xi|^2=10^{-3}$  
the signal significance after these cuts can reach 
to $S/\sqrt{B} \simeq 2.5$ and 
the signal over background ratio can be $S/B \sim 10$ \%, assuming 
the integrated luminosity of $1$ ab$^{-1}$.

\begin{figure}[t]
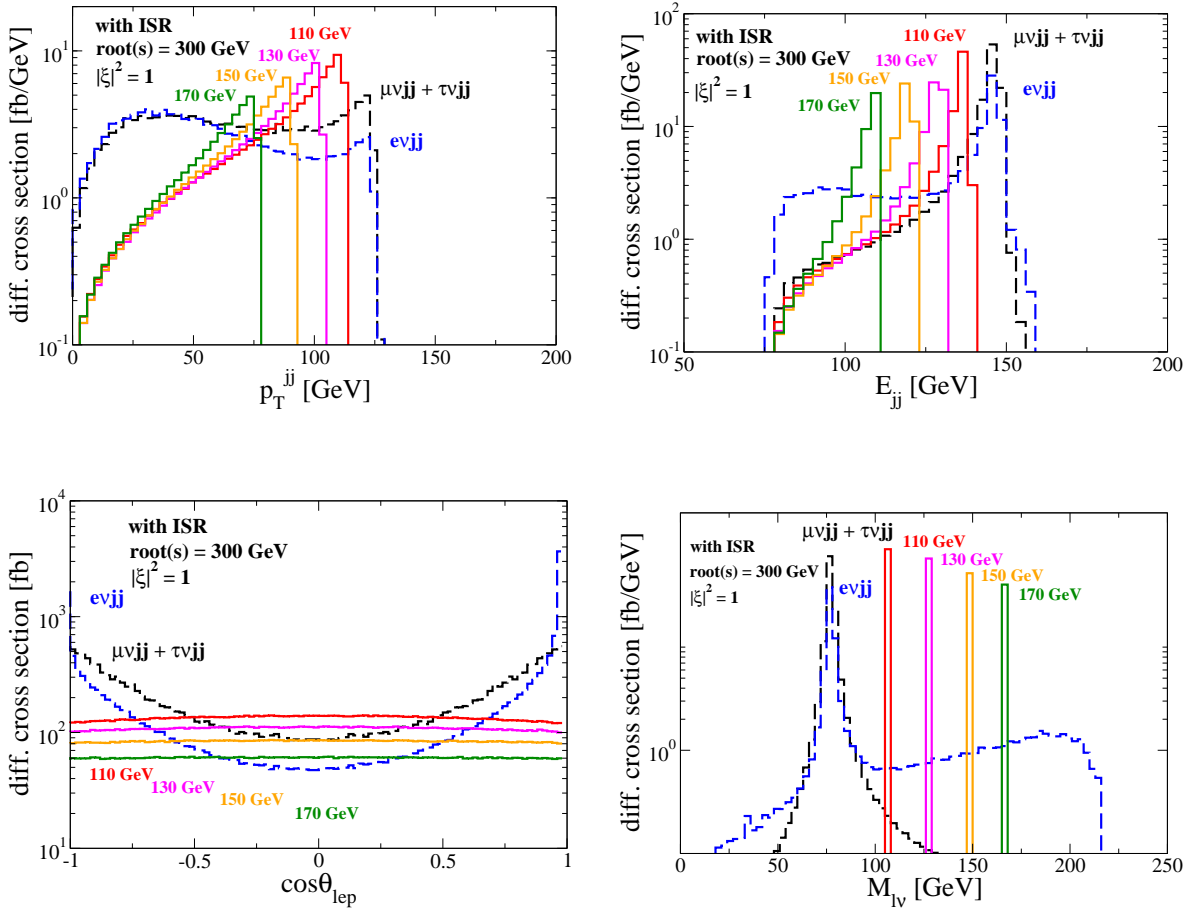

\begin{center}
\includegraphics[width=75mm]{distr_tjj_isr.eps}\hspace{5mm}
\includegraphics[width=75mm]{distr_ejj_isr.eps}\\
\vspace{10mm}
\includegraphics[width=75mm]{distr_clep_isr.eps}\hspace{5mm}
\includegraphics[width=75mm]{distr_mll_isr.eps}
\caption{Distributions of the signal for $m_{H^\pm}=110$, 
130, 150 and 170 GeV as well as the backgrounds 
after the invariant mass $M_{jj}$ cut in Eq.~(\ref{mjj}) with the ISR   
as a function of the transverse momentum 
$p_T^{jj}$ (upper left), the energy of the $jj$ system (upper right), 
the angle $\theta_{\rm lep}$ of a charged lepton with the beam 
axis (lower left), and 
the invariant mass $M_{\ell\nu}$ of the charged lepton and the missing momentum 
in the final state (lower right).  
$|\xi|^2$ is taken to be 1.}
\label{distr2}
\end{center}
\end{figure}
Next let us see how this results can be changed by including the ISR. 
The beam parameters at $\sqrt{s}=500$~GeV are given 
in Ref.~\cite{ILCTDR2007}, such as the bunch $x+y$ size, the bunch length 
and the number of particles per a bunch. 
We here use the default values defined in CalcHEP~\cite{CalcHEP}; i.e., 
the bunch $x+y$ size $=560$ nm, bunch length $=400$ 
$\mu$m, and the number of particles/bunch $=2 \times 10^{10}$ at 
$\sqrt{s}=300$~GeV\footnote{
We have confirmed that the results are almost unchanged 
even when we use the values given in Ref.~\cite{ILCTDR2007}.}.

\begin{figure}[t]
\begin{center}
\includegraphics[width=80mm]{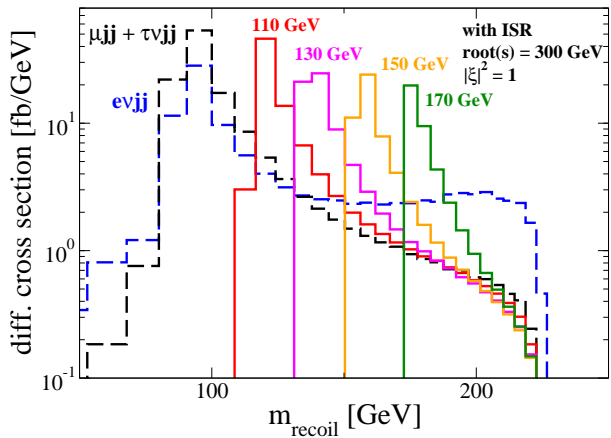}
\caption{Distributions of the signal for $m_{H^\pm}=110$, 130, 150 and 170 GeV
 as well as the backgrounds after the cut in Eq.~(\ref{mjj}) 
with the ISR as a function of the recoil mass $m_{\rm recoil}$.}
\label{distr_recoil_isr}
\end{center}
\end{figure}

In FIG.~\ref{distr2}, the similar distributions to those in 
FIG.~\ref{distr1} but with the ISR are given 
for the signal and the backgrounds after 
the invariant mass $M_{jj}$ cut in Eq.~(\ref{mjj}).  
The biggest change can be seen in the $E_{jj}$ distribution. 
The background events originally located at the point just below 150 GeV 
in the case without the ISR,  
which corresponds to the $W$ boson mass, 
tend to move in the lower $E_{jj}$ regions, 
so that the signal over background ratio becomes worse.  
The recoil mass distribution is shown in FIG.~\ref{distr_recoil_isr}. 

Consequently, the signal significance after all the cuts is smeared 
from $2.5$ to $2.0$, while the signal over background 
ratio is changed from $8.7\times 10^{-2}$ to $7.5 \times 10^{-2}$. 
Cross sections of the signal and the backgrounds with the ISR are 
listed in TABLE~\ref{result2} 
with the values of $S/\sqrt{B}$ and $S/B$ for each 
stage of kinematic cuts. 
We stress that even taking the ISR into account, 
the $H^\pm W^\mp Z$ vertex 
with $|\xi|^2 > 10^{-3}$ can be  excluded with 95\%~CL.

\section{Discussions}

In the above analysis, we have not distinguished differences  
among the charged leptons $\ell^\pm=e^\pm$, $\mu^\pm$ and $\tau^\pm$. 
Although we have assumed that the $H^\pm$ decay is 
lepton specific, the branching ratios of $H^\pm \to e^\pm \nu$, 
$\mu^\pm \nu$ and $\tau^\pm \nu$ depend on details of each 
Higgs model. 
For example, in the type~II seesaw model~\cite{type2seesaw}, 
the decay pattern of $H^\pm$ is related to the neutrino mass 
and mixing. Once we assume the decay pattern, 
we can easily estimate the signal significance 
by using TABLE~\ref{result1} or \ref{result2} for each specific model.  
If $H^\pm$ mainly decays into $e^\pm\nu$, the final state 
is $e\nu jj$. The backgrounds become 70 \% of all the $\ell\nu jj$ 
backgrounds as evaluated from TABLE~\ref{result2}, 
while the signal is unchanged.
The $S/\sqrt{B}$ becomes about 2.4 with taking into account the ISR.  
If $H^\pm$ mainly decays into $\mu^\pm\nu$, 
we consider the $\mu\nu jj$ final state. 
In this case, the number of the signal event 
is unchanged while that of the background becomes  
15 \% of all the $\ell\nu jj$ backgrounds, 
so that the $S/\sqrt{B}$ is expected to be multiplied 
by about 2.5 for the case with ISR; 
i.e., $S/\sqrt{B} \sim 5$ for $|\xi|^2 = 10^{-3}$. 
On the contrary, if $H^\pm$ mainly decays into $\tau^\pm\nu$, 
the $\tau\nu jj$ backgrounds 15 \% of all the $\ell\nu jj$ backgrounds. 
However, the efficiency for $\tau^\pm$ has to be 
multiplied so that it would  become worse than that for $\mu^\pm$.  

In extended Higgs sectors, there can be additional 
backgrounds which is relevant to the other scalar 
bosons than $H^\pm$. 
For example, if one of the neutral Higgs bosons $H$ has the 
similar mass to that of $H^\pm$, then the process 
$e^+e^- \to ZH \to \ell\ell jj$ (or $jj \ell\ell$) 
can be the background. In this case, 
its contribution is expected to be reduced by the kinematic 
cut of $M_{jj}$ in Eq.~(\ref{mjj}) with the good resolution 
of the ILC ($\sigma_E \sim 0.3 \sqrt{E}$ GeV). 
Also, the rate of missing one of the charged leptons 
has to be multiplied. 
The change of signal significance cannot be expected to 
be dominant.
As the second example, we mention about pair production of 
doubly charged Higgs bosons 
$e^+e^- \to H^{++} H^{--} \to \ell^+\ell^+ \ell^-\ell^-$ 
with the similar mass to that of $H^\pm$. 
Such degenerate masses between $H^\pm$ and $H^{\pm\pm}$ 
may often happen in the models with triplets. 
When the final state is of four charged leptons, 
we can reduce such a background by imposing 
the veto for three or four leptons in the final state.

Finally, we comment on the case where $H^\pm$ is relatively 
heavy as compared to the value we have assumed above.
When $m_{H^\pm} > m_W + m_Z$, the main decay mode of $H^\pm$ 
may be $WZ$, instead of $\ell \nu$.  
The final state of the signal would be 
$\ell \nu jjjj$, $\ell\ell \nu \nu jj$ or $\ell\ell jjjj$, so that 
the analysis should be different from the $\ell \nu jj$ 
final state. It is expected that the signal significance 
for such a case would become rather worse.  
Furthermore, when $m_{H^\pm} > 2 m_t$, the 
$t \overline{t}$ pair production mode becomes additional 
background.  
In order to clarify the feasibility of the $H^\pm W^\mp Z$ vertex 
for such a case, we need to proceed to the analysis for such 
final states, but this is beyond the scope of our paper.

\section{Conclusions}

We have discussed the possibility of measuring  the $H^\pm W^\mp Z$ vertex 
at the ILC. 
The vertex is important to understand the exoticness of the Higgs sector, 
so that the combined information of this vertex with the rho parameter 
provides a useful criterion to determine the structure of the extended Higgs sector. 
Assuming that the decay of the charged Higgs bosons is lepton specific, 
which is natural for the exotic representations of the extra scalar bosons, 
the feasibility of the vertex is analyzed by using the recoil method via the process 
$e^+e^- \to H^\pm W^\mp \to \ell\nu jj$ 
with the parton level simulation for the background reduction. 
We have found that the vertex with  $|F_{HWZ}|^2 \geq {\cal O}(10^{-3})$ can be 
excluded  with the 95\% confidence level when $120$-$130$ GeV $ < m_{H^\pm} < m_W+m_Z$.  
For heavier charged Higgs bosons, the decay into $WZ$ may be dominant so that 
the analysis becomes model dependent.  
The meaurement of the $H^\pm W^\mp Z$ vertex with 
$|F_{HWZ}|^2 \geq {\cal O}(10^{-3})$ gives a precise information 
for the Higgs sector, whose accuracy is similar to that of the rho parameter.
In conclusion, the measurement of the $H^\pm W^\mp Z$ vertex at the ILC is therefore 
very interesting, and give a motivation to perform the more realistic 
detector level simulation.

\vspace{5mm}
\noindent
{\it Acknowledgments}

The authors would like to thank Hitoshi Yamamoto and Keisuke Fujii 
for useful discussions and comments. 
The work of SK was supported in part by 
Grant-in-Aid for Scientific Research (A) no.
22244031 and (C) no. 19540277, 
that of KY was supported by Japan Society for the 
Promotion of Science (JSPS Fellow (DC2)).


\begin{thebibliography}{1}

\bibitem{rho_exp}
 
K.~Nakamura, et al., (Particle Data Group), 
J. Phys. G {\bf 37}, 075021 (2010). 


\bibitem{rho_formula}

  J.~F.~Gunion, H.~E.~Haber, G.~L.~Kane and S.~Dawson,
  ``THE HIGGS HUNTER'S GUIDE,''
  Front.\ Phys.\  {\bf 80}, 1 (2000).

\bibitem{rho}
  E.~Gildener and S.~Weinberg,
  Phys.\ Rev.\  D {\bf 13}, 3333 (1976).

\bibitem{rho_loop}

  J.~Alcaraz {\it et al.} [ ALEPH and DELPHI and L3 and OPAL and LEP Electroweak Working Group Collaborations ],
[hep-ex/0612034].
 

\bibitem{STU}

  M.~E.~Peskin and T.~Takeuchi,
  Phys.\ Rev.\ Lett.\  {\bf 65}, 964 (1990); 
  Phys.\ Rev.\  D {\bf 46}, 381 (1992).

\bibitem{Grifols-Mendez}
  J.~A.~Grifols and A.~Mendez,
  Phys.\ Rev.\  D {\bf 22}, 1725 (1980); 
A.A.~Iogansen, N.G.~Ural\'{t}sev, V.A.~Khoze, 
{\it Sov. J. Nucl. Phys.} {\bf 36} {(1982)} {717}.

\bibitem{HWZ}
  A.~Mendez and A.~Pomarol,
  Nucl.\ Phys.\  B {\bf 349}, 369 (1991);
  M.~C.~Peyranere, H.~E.~Haber and P.~Irulegui,
  Phys.\ Rev.\  D {\bf 44}, 191 (1991);
J.L.~D\'{i}az-Cruz, J.~Hern\'{a}ndez-S\'{a}nchez, J.J.~Toscano,
  Phys.\ Lett.\ B {\bf512}, 339 (2001).

\bibitem{HWZ-Kanemura}
  S.~Kanemura,
  Phys.\ Rev.\  D {\bf 61}, 095001 (2000). 

\bibitem{logan} H.~Haber, H.~Logan, Phys.\ Rev.\ D {\bf 62}, 015011 (2000).

\bibitem{Glashow:1976nt}
  S.~L.~Glashow and S.~Weinberg,
  Phys.\ Rev.\  D {\bf 15}, 1958 (1977).

\bibitem{thdm_Yukawa1}

  V.~D.~Barger, J.~L.~Hewett and R.~J.~N.~Phillips,
  Phys.\ Rev.\  D {\bf 41}, 3421 (1990);
  Y.~Grossman,
  Nucl.\ Phys.\  B {\bf 426}, 355 (1994).

\bibitem{thdm_Yukawa2}

  M.~Aoki, S.~Kanemura, K.~Tsumura and K.~Yagyu,
  Phys.\ Rev.\  D {\bf 80}, 015017 (2009); 
  H.~S.~Goh, L.~J.~Hall and P.~Kumar,
  JHEP {\bf 0905}, 097 (2009);
%
  S.~Su and B.~Thomas,
  Phys.\ Rev.\  D {\bf 79}, 095014 (2009);
%
  H.~E.~Logan and D.~MacLennan,
  Phys.\ Rev.\  D {\bf 79}, 115022 (2009).

\bibitem{Georgi:1985nv}
  H.~Georgi and M.~Machacek,
  Nucl.\ Phys.\  B {\bf 262}, 463 (1985); 
  M.~S.~Chanowitz and M.~Golden,
  Phys.\ Lett.\  B {\bf 165}, 105 (1985).

\bibitem{Gunion:1989ci}
  J.~F.~Gunion, R.~Vega and J.~Wudka,
  Phys.\ Rev.\  D {\bf 42}, 1673 (1990); 
  R.~Vega and D.~A.~Dicus,
  Nucl.\ Phys.\  B {\bf 329}, 533 (1990).
  J.~F.~Gunion, R.~Vega and J.~Wudka,
  Phys.\ Rev.\  D {\bf 43}, 2322 (1991); 
  R.~Godbole, B.~Mukhopadhyaya and M.~Nowakowski,
  Phys.\ Lett.\  B {\bf 352}, 388 (1995).

\bibitem{AK_GM} 
    
  M.~Aoki and S.~Kanemura,
  Phys.\ Rev.\  D {\bf 77}, 095009 (2008); 
  H.~E.~Logan, M.~-A.~Roy,
  Phys.\ Rev.\  {\bf D82}, 115011 (2010).
  [arXiv:1008.4869 [hep-ph]].

\bibitem{topdecay}

A.~C.~Bawa, C.~S.~Kim and A.~D.~Martin,
Z.\ Phys.\ C {\bf 47} (1990) 75;

\bibitem{SingleH+}

  J.~F.~Gunion, H.~E.~Haber, F.~E.~Paige, W.~K.~Tung and S.~S.~D.~Willenbrock,
  Nucl.\ Phys.\  B {\bf 294}, 621 (1987).


\bibitem{ggWH+} 

D.~A.~Dicus, J.~L.~Hewett, C.~Kao and T.~G.~Rizzo,
Phys.\ Rev.\ D {\bf 40} (1989) 787; 
A.~A.~Barrientos Bendez\'{u} and B.~A.~Kniehl,
Phys.\ Rev.\ D {\bf 59} (1999) 015009;
ibid.\ D {\bf 61} (2000) 097701;
ibid.\ D {\bf 63} (2001) 015009;
O.~Brein, W.~Hollik and S.~Kanemura,
Phys.\ Rev.\ D {\bf 63} (2001) 095001; 
Y.~S.~Yang, C.~S.~Li, L.~G.~Jin and S.~H.~Zhu,
Phys.\ Rev.\ D {\bf 62} (2000) 095012; 
F.~Zhou, W.~G.~Ma, Y.~Jiang, L.~Han and L.~H.~Wan,
Phys.\ Rev.\ D {\bf 63} (2001) 015002;
W.~Hollik and S.~H.~Zhu,
Phys.\ Rev.\ D {\bf 65} (2002) 075015; 
  E.~Asakawa, O.~Brein and S.~Kanemura,
  Phys.\ Rev.\  D {\bf 72}, 055017 (2005);
  D.~Eriksson, S.~Hesselbach, J.~Rathsman,
  Eur.\ Phys.\ J.\  {\bf C53}, 267-280 (2008).

\bibitem{ggHpHm}

S.~Willenbrock, Phys.\ Rev.\ D {\bf 35}, 173 (1987);
O.~Brein and W.~Hollik, Eur.\ Phys.\ J. C {\bf 13}, 175 (2000);
A.A.~Barrientos Bendezu and B.A.~Kniehl, Phys.\ Rev.\ D {\bf 64}, 035006 (2001).

\bibitem{qqPairH+} 

  E.~Eichten, I.~Hinchliffe, K.~D.~Lane and C.~Quigg,
  Rev.\ Mod.\ Phys.\  {\bf 56}, 579 (1984)
  [Addendum-ibid.\  {\bf 58}, 1065 (1986)].

\bibitem{AH+} 
  S.~Kanemura and C.~P.~Yuan,
  Phys.\ Lett.\  B {\bf 530}, 188 (2002); 
  Q.~H.~Cao, S.~Kanemura and C.~P.~Yuan,
  Phys.\ Rev.\  D {\bf 69}, 075008 (2004);
  A.~Belyaev, Q.~-H.~Cao, D.~Nomura, K.~Tobe, C.~-P.~Yuan,
  Phys.\ Rev.\ Lett.\  {\bf 100}, 061801 (2008).

\bibitem{H-H++} 

  A.~G.~Akeroyd, M.~Aoki,
  Phys.\ Rev.\  {\bf D72}, 035011 (2005). 

\bibitem{WZfusion}

  E.~Asakawa and S.~Kanemura,
  Phys.\ Lett.\  B {\bf 626}, 111 (2005);
  E.~Asakawa, S.~Kanemura and J.~Kanzaki,
  Phys.\ Rev.\  D {\bf 75}, 075022 (2007);
  M.~Battaglia, A.~Ferrari, A.~Kiiskinen, T.~Maki,
  [hep-ex/0112015];
  S.~Godfrey, K.~Moats,
  Phys.\ Rev.\  {\bf D81}, 075026 (2010).


\bibitem{ILCTDR2007}
 
  J.~Brau, (Ed.) {\it et al.} [ ILC Collaboration ],
    [arXiv:0712.1950 [physics.acc-ph]]; \\
  G.~Aarons {\it et al.} [ ILC Collaboration ],
  [arXiv:0709.1893 [hep-ph]]; \\
  T.~Behnke, (Ed.) {\it et al.} [ ILC Collaboration ],
  [arXiv:0712.2356 [physics.ins-det]].

\bibitem{ILC_PairH+}

  S.~Komamiya,
  Phys.\ Rev.\  D {\bf 38}, 2158 (1988);
  A.~Djouadi, J.~Kalinowski, P.~Ohmann, P.~M.~Zerwas,
  Z.\ Phys.\  {\bf C74}, 93-111 (1997);
  A.~Kiiskinen, P.~Poyhonen, M.~Battaglia; 
  [hep-ph/0101239]; 
  J.~Guasch, W.~Hollik, A.~Kraft,
  Nucl.\ Phys.\  {\bf B596}, 66-80 (2001).

\bibitem{Gamma_PairH+}

  D.~Bowser-Chao, K.~-m.~Cheung, S.~D.~Thomas,
  Phys.\ Lett.\  {\bf B315}, 399-405 (1993).

\bibitem{ILC_ffH+}

  S.~Kanemura, S.~Moretti and K.~Odagiri,
  JHEP {\bf 0102}, 011 (2001).

\bibitem{Cheung:1994rp}
  K.~Cheung, R.~J.~N.~Phillips and A.~Pilaftsis,
  Phys.\ Rev.\  D {\bf 51}, 4731 (1995).

\bibitem{eeHW1}

  S.~Kanemura,
  Eur.\ Phys.\ J.\  C {\bf 17}, 473 (2000).

\bibitem{eeHW2}
S.-H.~Zhu, hep-ph/9901221;
A. Arhrib, {et al.}, Nucl.\ Phys.\ B {\bf 581}, 34 (2000);
%
H.E.~Logan, S.~Su, Phys.\ Rev.\ D {\bf 66}, 035001 (2002);
Phys.\ Rev.\ D {\bf 67}, 017703, (2003); 
%
K.~Cheung, R.~Phillips, A.~Pilaftsis, 
Phys.\ Rev.\ D {\bf 51}, 4731 (1995);
R.M.~Godbole, B.~Mukhopadhyaya, M.~Nowakowski,
Phys.\ Rev.\ D {\bf 352}, 388 (1995);
D.K.~Ghosh, R.M.~Godbole, B.~Mukhopadhyaya,
Phys.\ Rev.\ D {\bf 55}, 3150 (1997).

\bibitem{Gamma_SingleH+}

  H.~J.~He, S.~Kanemura and C.~P.~Yuan,
  Phys.\ Rev.\ Lett.\  {\bf 89}, 101803 (2002);
  H.~J.~He, S.~Kanemura and C.~P.~Yuan,
  Phys.\ Rev.\  D {\bf 68}, 075010 (2003);
  S.~Moretti and S.~Kanemura,
  Eur.\ Phys.\ J.\  C {\bf 29}, 19 (2003).

\bibitem{eGamma_SingleH+}

  S.~Kanemura, S.~Moretti and K.~Odagiri,
  Eur.\ Phys.\ J.\  C {\bf 22}, 401 (2001).
  

\bibitem{ZH}

  J.~R.~Ellis, M.~K.~Gaillard, D.~V.~Nanopoulos,
  Nucl.\ Phys.\  {\bf B106}, 292 (1976); 
J.~D.~Bjorken, SLAC Report, 198 (1976); 
  B.~L.~Ioffe, V.~A.~Khoze,
  Sov.\ J.\ Part.\ Nucl.\  {\bf 9}, 50 (1978);  
  D.~R.~T.~Jones, S.~T.~Petcov,
  Phys.\ Lett.\  {\bf B84}, 440 (1979).

\bibitem{recoil}

  W.~Lohmann, M.~Ohlerich, A.~Raspereza and A.~Schalicke,
{\it In the Proceedings of 2007 International Linear Collider Workshop (LCWS07 and ILC07), Hamburg, Germany, 30 May - 3 Jun 2007, pp TIG13}
  [arXiv:0710.2602 [hep-ex]];
  H.~Li, F.~Richard, R.~Poeschl and Z.~Zhang,
  arXiv:0901.4893 [hep-ex].

\bibitem{CalcHEP}
  A.~Pukhov,
  [hep-ph/0412191].

\bibitem{Kanemura:1993hm}
  S.~Kanemura, T.~Kubota and E.~Takasugi,
  Phys.\ Lett.\  B {\bf 313}, 155 (1993); 
  A.~G.~Akeroyd, A.~Arhrib, E.~-M.~Naimi,
  Phys.\ Lett.\  {\bf B490}, 119-124 (2000).
  I.~F.~Ginzburg and I.~P.~Ivanov,
  Phys.\ Rev.\  D {\bf 72}, 115010 (2005).

\bibitem{Haber-Pomarol} 

  H.~E.~Haber, A.~Pomarol,
  Phys.\ Lett.\  {\bf B302}, 435-441 (1993); 
  A.~Pomarol, R.~Vega,
  Nucl.\ Phys.\  {\bf B413}, 3-15 (1994).

\bibitem{type2seesaw} 

J.~Schechter and J.~W.~F.~Valle,
 Phys.\ Rev.\  D {\bf 22} (1980) 2227;
T.~P.~Cheng and L.~F.~Li,
 Phys.\ Rev.\  D {\bf 22} (1980) 2860;
M.~Magg and C.~Wetterich,
 Phys.\ Lett.\  B {\bf 94} (1980) 61;
C.~Wetterich,
 Nucl.\ Phys.\  B {\bf 187} (1981) 343;
G.~Lazarides, Q.~Shafi and C.~Wetterich,
 Nucl.\ Phys.\  B {\bf 181} (1981) 287;
R.~N.~Mohapatra and G.~Senjanovic,
 Phys.\ Rev.\  D {\bf 23} (1981) 165.


\end{thebibliography}
\end{document}